# Extending the SDSS Batch Query System to the National Virtual Observatory Grid

María A. Nieto-Santisteban,
William O'Mullane
Nolan Li
Tamás Budavári
Alexander S. Szalay
Aniruddha R. Thakar
Johns Hopkins University

Jim Gray
Microsoft Research

# Extending the SDSS Batch Query System
# to the National Virtual Observatory Grid

María A. Nieto-Santisteban[1], William O'Mullane[1], Jim Gray[2], Nolan Li[1],
Tamás Budavári[1], Alexander S. Szalay[1], Aniruddha R. Thakar[1]

**Abstract**. The Sloan Digital Sky Survey science database is approaching 2TB. While the vast majority of queries normally execute in seconds or minutes, this interactive execution time can be disproportionately increased by a small fraction of queries that take hours or days to run; either because they require non-index scans of the largest tables or because they request very large result sets. In response to this, we added a multi-queue job submission and tracking system. The transfer of very large result sets from queries over the network is another serious problem. Statistics suggested that much of this data transfer is unnecessary; users would prefer to store results locally in order to allow further cross matching and filtering. To allow local analysis, we implemented a system that gives users their own personal database (MyDB) at the portal site. Users may transfer data to their MyDB, and then perform further analysis before extracting it to their own machine.

We intend to extend the MyDB and asynchronous query ideas to multiple NVO nodes. This implies development, in a distributed manner, of several features, which have been demonstrated for a single node in the SDSS Batch Query System (CasJobs). The generalization of asynchronous queries necessitates some form of MyDB storage as well as workflow tracking services on each node and coordination strategies among nodes.

## 1. Sloan Digital Sky Survey – SkyServer

Web access to the SDSS databases via the SkyServer[3] has been publicly available since June 2001. The SkyServer front end is coded in ASP on a Microsoft.Net server and backed by a SQL Server database. The current database, Data Release 1 (DR1), is over 1TB (with indexes), the release in preparation is 2TB, and subsequent releases will bring this to at least 6TB of catalog data in SQL Server. In fact there will be up to 50 TB of pixel and catalog data and more of this may be put in the database. For example, the data points for all SDSS[4] spectra were recently loaded into a separate database. Hence the database could become much larger than 6TB. The SkyServer site allows interactive queries in SQL[5]. The results of some of these queries are large, averaging millions of rows. The site averages 2M hits per month. Considering it is running on a $10k server, the site performs extremely well. However in certain circumstances we have experienced problems. Complex queries can swamp the system and erroneous queries may run for a long time but never complete.

### 1.1. SkyServer Statistics

Figure 1 indicates that the SkyServer execution times and result set sizes follow a natural power law. Hence there is no particularly obvious point at which queries could be cut off. All queries currently run at the same priority - there is no effective query governor or priority scheduling system built into SQL Server. While this may not be a problem in itself, long queries can slow down the system, causing what should be quick queries to take much longer. Some long queries spend most of their time returning large result sets to a user over the Internet. We have seen as many as twelve million rows (20GB) downloaded in one hour and single result sets typically of 1M rows each. These large transfers are often unnecessary; this data is often used only to make comparisons against a small local set or is used in the next stage of an interactive query session exploring the data.

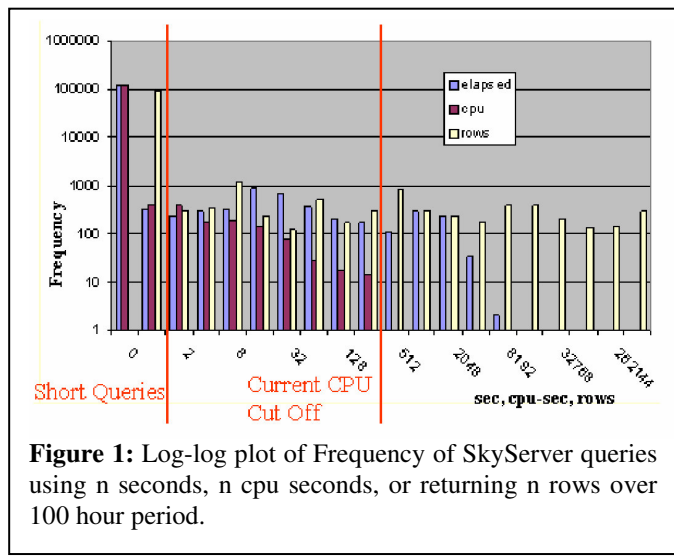

**Figure 1:** Log-log plot of Frequency of SkyServer queries using n seconds, n cpu seconds, or returning n rows over 100 hour period.

---

[1] The Johns Hopkins University
[2] Microsoft Research
[3] http://skyserver.sdss.org
[4] Sloan Digital Sky Survey
[5] Structured Query Language

## 2. Batch System

We developed a simple batch processing system[6] to address these problems. It provides query distribution by size, allows rudimentary load balancing across multiple machines, guarantees query completion or termination, provides local storage for users, and separates database query from data extraction and download. This will be pertinent for the Virtual Observatory as the SkyNode protocol begins to mature (Yasuda et al. 2004, Budavári et al. 2004).

### 2.1. Queues

We have multiple queues based on query length. Jobs in the shortest jobs queue are processed as soon as possible on a first come, first serve basis. Jobs in all other queues are redirected and executed sequentially (limited concurrent execution is allowed) on different machines each mirroring a separate copy of the data. Query execution time is strictly limited by the limit assigned to a particular queue. A job may take only as long as the limit of its queue or it will be terminated. Hence there are no ghost jobs hanging around for days, nor can a long query hinder execution of a shorter one.

### 2.2. Query Estimation

There is a query estimator in SQL server however its accuracy is questionable and for the first version of this system we decided not to guess query execution time. This responsibility is left to the users; they decide how long they think their queries will take and choose the appropriate queue accordingly. As mentioned previously, queries exceeding queue time limit will be canceled. We do provide *autocomplete* functionality that moves a query to the next queue if it times out in its original queue. In a future release we may use the statistics gathered on jobs to develop a heuristic for estimating query lengths and automatically assigning them to queues.

### 2.3. MyDB - Local Storage

Queries submitted to the longer queues must write results to local storage, known as MyDB, using the standard SQL select-into syntax e.g.

```
select top 10 *
into   MyDB.rgal
from   galaxy
where  r < 22 and r >21
```

The MyDB idea is similar to the AstroGrid MySpace (Walton et al. 2004) notion. We create a SQL Server database for the user dynamically the first time MyDB is used in a query. Upon creation, appropriate database links and grants are made such that the database will work in queries on any database. Since this is a normal database the user may perform joins and queries on tables in MyDB as with tables in any other database. The user is responsible for this space and may drop tables from it to keep it clear. We initially assign each user 100MB but this is configurable on a system and per user basis.

### 2.4. Groups

Some users may wish to share data in their MyDBs. Any user with appropriate privileges may *create* a group and *invite* other users to *join* the group. An invited user may *accept* group membership. A user may then publish any of his MyDB tables to the groups of which he is a member. Other group members may access these tables by using a pseudo database name consisting of the word *group* followed by the *id* of the other user followed by the table name. For example, if the c*osmology* user published the table *rgal* and if you were a *cosmology* group member, you could access this table using the name *GROUP.cosmology.rgal* in your queries.

---

[6] http://skyservice.pha.jhu.edu/devel/casjobs

## 2.5. Import/Export Tables

Tables from MyDB may be requested in FITS (Hanisch et al. 2001), CSV, or VOTable[7] format. Extraction requests are queued as a different job type and have their own processor. File extraction is done on the server. A URL to the file is put in the job record upon completion. Group tables also appear in the extraction list. Users may also upload a CSV file of data to an existing table in MyDB. Having the table created before upload gives appropriate column names right type conversions, for examples integers, floats and dates are recognized. Without this schema information all data would be treated as strings and the table would have column names like col5. We hope the ability to upload data and the group system will reduce some of the huge downloads from our server.

## 2.6. Jobs

Apart from the short jobs, everything in this system is asynchronous and requires job tracking. The tracking is done by creating and updating a row in a Jobs table in the administrative database. However this also requires users to be identified and associated with the jobs. Identification is also required for authoring and resolving access to MyDB. Representing jobs in a database makes it easy to implement management, scheduling, and query operations. A user may list all previous jobs and get the details of status, time submitted, started, finished, etc. The user may also resubmit a job.

## 2.7. Ferris Wheel

A future experiment we will try is to batch full table scan queries together. We will piggy-back queries in SQL Server so that a single sequential scan is made of the data instead of several. Ideally we would like to not have to wait for a set of queries to finish scanning to join the batch. Rather we would like some set of predefined entry points where a new query could be added to the scan. Conceptually one may think of this as a Ferris wheel where no matter which bucket you enter you will be given one entire revolution.

## 3. SOAP Services

We found that SOAP and WSDL Web services provide a very clean API for any system. In this system the Web site sits upon a set of SOAP-WSDL services. Any user may access these services directly using a SOAP toolkit in their preferred programming language. Most of our clients are in C# and JavaScript, but we have successfully used Python and Java (AXIS) clients for Web services. Others have written Perl clients. More information on this is available at the IVOA Web site[8].

## 4. MyDB for NVO

We are extending the MyDB and asynchronous query ideas to multiple NVO nodes. This implies development of multi-node generalization of the CasJobs features. The introduction of asynchronous queries necessitates some form of MyDB storage as well as workflow tracking facilities.

We will specify a generic SOAP-WSDL interface for these tasks. This will allow heterogeneous data and job engines to have a common interface for submission and querying of jobs. CasJobs shows that providing local space on a node is trivial. Still even in that trivial case we needed security and naming mechanisms to allow selective sharing. However, distributed job execution requires user identification/authentication beyond simple passwords. We want to be authenticated once and then not challenged at further nodes. The nodes would of course still perform their own authorization. Hence some form of distributed authentication and certificate system is required. Such systems are widely deployed in the commercial space (e.g. Passport, Liberty), and others are under development in the Grid and the Web services communities – these seem to finally be converging, making it an opportune moment for us to exploit this technology.

We are exploring a distributed system whereby multiple nodes form a trusted network (Virtual Organization) and users with an account (i.e. authentication) on one node will be accepted at other nodes without needing to register. Interaction with other initiatives similar to MyDB, such as AstroGrid's MYSPACE will allow us to experiment with issues such as how exactly the Virtual Organization will work i.e. *if I am allowed MyDB by NVO should AstroGrid*

---

[7] http://www.us-vo.org/VOTable/
[8] http://www.ivoa.net/twiki/bin/view/IVOA/WebgridTutorial

*allow MYSPACE?* We foresee then a user having datasets in multiple "spaces" on multiple machines. We are investigating how to build a portal and appropriate protocols to enable the display of what data and jobs a given user has on which machines and the status of the workflow.

Some other interesting issues we need to explore are how to automatically join data and results from different spaces and bring them into MyDB, and how to validate the data and track the workflow. SDSS has already in place an automatic loading system, the sqlLoader, which imports data stored in CSV files into the SkyServer database (Thakar, Szalay, Gray 2004).

### 4.1. sqlLoader

The sqlLoader is a distributed workflow system of SQL modules that check, load, validate and publish the data to the SDSS databases. The workflow is described by a directed acyclic graph whose nodes are the processing modules. Each node has an *undo* compensation step in case the job has to be canceled. It is designed for parallel loading and is controlled from a Web interface (Load Monitor). The validation stage represents a systematic and thorough scrubbing of the data. The different data products are merged into a set of linked tables that can be efficiently searched with specialized indices and pre-computed joins. Although the system currently collects the data from files stored locally, we could easily make it generic and collect the data from any given space as long as the nodes provide the necessary data delivery services. Other than that, all the validation and workflow tracking steps would remain the same. This would be an excellent test case for Yukon, the next version of Sql Server, which will provide native Web services support in the database. The sqlLoader could consume the data directly from the Web services provided by the SkyNodes.

### 5. Summary

The CasJobs system provides a good test bed for a single-node Virtual Observatory. It provides job queuing and scheduling, it provides local data storage for users that reduces network traffic and speeds query execution. It provides a primitive security system of groups and selective sharing.

Each of these features will be very useful in the more general context of a multi-archive Virtual Observatory portal spread among multiple geographically distributed nodes. We are in the process of designing and implementing these generalizations.